\documentclass[conference]{IEEEtran}
\ifCLASSINFOpdf
\else
\fi
\usepackage{multirow}
\usepackage{cite}
\usepackage{amsmath,amssymb,amsfonts}
\usepackage{graphicx}
\usepackage{textcomp}
\usepackage{xcolor}
\usepackage{tabularx}
\usepackage{makecell}
\def\BibTeX{{\rm B\kern-.05em{\sc i\kern-.025em b}\kern-.08em
    T\kern-.1667em\lower.7ex\hbox{E}\kern-.125emX}}

\usepackage{booktabs}
\usepackage{algpseudocode}
\usepackage{algorithm}
\usepackage{amsmath}
\usepackage{amssymb}
\usepackage{multirow}
\usepackage{subcaption}
\usepackage{hyperref}
\usepackage{ulem} 
\usepackage{tikz} 
\usepackage{mathptmx} 
\usepackage{float}
\usepackage{graphicx}
\usepackage{subcaption}

\usepackage{fancyhdr}
\usepackage{lipsum}  

\fancypagestyle{firstpage}{
  \fancyhf{} 
  \fancyhead[L]{IEEE International Conference on Wearable and Implantable Body Sensor Networks (BSN 2024)} 
  \fancyfoot[C]{\thepage} 
}

\begin{document}
%
\title{MotionTrace: IMU-based Field of View Prediction for Smartphone AR Interactions}

\author{
\IEEEauthorblockN{
Rahul Islam \href{https://orcid.org/0000-0003-3601-0078}{\includegraphics[scale=0.06]{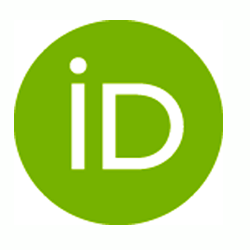}}
}
\IEEEauthorblockA{
Stevens Institute of Technology\\
Hoboken, USA\\
}

\and

\IEEEauthorblockN{
Vasco Xu \href{https://orcid.org/0000-0003-3990-582X}{\includegraphics[scale=0.06]{Fig/orcid.png}}
}
\IEEEauthorblockA{
University of Chicago\\
Chicago, USA\\
}

\and

\IEEEauthorblockN{
Karan Ahuja
\href{https://orcid.org/0000-0003-2497-0530}{\includegraphics[scale=0.06]{Fig/orcid.png}}
}
\IEEEauthorblockA{
Northwestern University\\
Evanston, USA\\
}

}


%


\maketitle
\thispagestyle{firstpage}

\begin{abstract}
For handheld smartphone AR interactions, bandwidth is a critical constraint. Streaming techniques have been developed to provide a seamless and high-quality user experience despite these challenges. To optimize streaming performance in smartphone-based AR, accurate prediction of the user's field of view is essential. This prediction allows the system to prioritize loading digital content that the user is likely to engage with, enhancing the overall interactivity and immersion of the AR experience. In this paper, we present MotionTrace, a method for predicting the user's field of view using a smartphone's inertial sensor. This method continuously estimates the user's hand position in 3D-space to localize the phone position. We evaluated MotionTrace over future hand positions at 50, 100, 200, 400, and 800ms time horizons using the large motion capture (AMASS) and smartphone-based full-body pose estimation (Pose-on-the-Go) datasets. We found that our method can estimate the future phone position of the user with an average MSE between 0.11 - 143.62 mm across different time horizons.
\end{abstract}


%
\IEEEpeerreviewmaketitle

\section{Introduction}

Augmented Reality (AR) blends digital elements with the real world, creating interactive experiences in sectors like healthcare, education, and entertainment. With tools such as Apple's ARKit and Google's ARCore, AR technology is readily available on everyday devices like smartphones and tablets. By 2024, AR has reached a billion users. AR experiences enhance the user's field of view (FOV) by integrating digital content, enabling interaction with digital elements overlaid on physical surroundings. However, the widespread adoption of AR is hindered by the need for high bandwidth and continuous tracking for realistic experiences. High-quality AR experiences require significant data transfers, including complex meshes and textures, to make digital objects appear realistic within the user's FOV. This results in extensive rendering and possible delays, leading to subpar user experiences due to prolonged loading times before an AR experience can start.

Field-of-View (FOV)-dependent streaming, initially developed for 360-degree video platforms, optimizes video delivery by adapting the streaming quality to the user's current and projected FOV, significantly reducing startup latency and data transmission volume \cite{bao2016shooting, sun2018multi}. This concept has been extended to augmented reality (AR) applications, where techniques selectively enhance resolution and detail of AR content within or likely to fall within the user's immediate view, improving performance and user experience \cite{wang2020towards}. However, AR presents unique challenges due to the complex and dynamic nature of its environments, where multiple objects often occupy the user's FOV simultaneously, necessitating a more comprehensive approach to FOV prediction and rendering of immediately relevant AR content with higher quality. Continuous camera-based tracking in AR can degrade smartphone performance due to high power consumption and processing demands, potentially leading to overheating and reduced battery life, which in turn negatively impacts user experience by limiting the device's operational duration and responsiveness. Despite initial successes, the scope of FOV prediction in AR remains under-explored, offering avenues for significant improvements and innovations.

FOV prediction in AR, distinguished from 360-degree video streaming by AR's support for six degrees of freedom (6DOF) and interactions with digitally superimposed objects on the real world, remains an unresolved problem requiring tailored solutions \cite{bao2016shooting, petrangeli2018trajectory, viola2021trace}. We propose MotionTrace, a novel approach that utilizes the readily accessible Inertial Measurement Unit (IMU) data from smartphones—leveraging historical data on hand position combined with orientation and acceleration—to accurately predict user’s FOV in AR, focusing solely on translation across three degrees of freedom. This method not only addresses the complexity added by AR’s interactive digital objects and user movements but also highlights the superiority of using IMU over camera sensors. IMU-based tracking is less resource-intensive, making it more suitable for continuous operation without substantial power drain, thus ideal for long-term applications where battery conservation is critical. Additionally, the IMU proves reliable where camera-based systems falter, such as in poorly lit or visually occluded environments. This makes IMUs exceptionally beneficial for persistent sensing in diverse conditions. Our extensive evaluation of this method uses large motion capture (AMASS) and smartphone-based full-body pose estimation (Pose-on-the-Go) datasets, effectively predicting future hand positions at intervals up to 800ms at 30fps, showcasing the practical applicability and advantages of IMU data in dynamic, interactive AR settings \cite{viola2021trace}.
\section{Related Work}

\subsection{Field of View Prediction in Augmented Reality}

In bandwidth-demanding AR applications, several methods have been developed to optimize the streaming of AR content, enhancing user experience and reducing bandwidth. Noh et al. \cite{noh2020cloud} and Park et al. \cite{park2018volumetric} describe cloud-assisted systems and 3D tiling techniques, respectively, that select optimal levels of detail and tiles based on bandwidth and user proximity. Crucial to these technologies is the accurate prediction of a user’s field of view (FOV), which is enhanced by algorithms like the Trace Match \& Merge \cite{viola2021trace}, using historical AR data to predict future FOV, and the ACE Dataset approach \cite{wang2020towards}, which analyzes user movements and digital object locations to predict user focus areas. These predictive models are essential for maintaining high visual fidelity and surpass traditional FOV prediction methods such as dead reckoning and linear regression \cite{bao2016shooting}.

By predicting which parts of a scene a user is likely to focus on, these systems can pre-load high-fidelity graphics in those areas, thereby reducing latency and enhancing the overall user experience. Such techniques underscore the importance of predictive accuracy in the development of advanced AR platforms, providing a direct link to the necessity of our research in improving FOV prediction through MotionTrace.

\subsection{Inertial Sensors in Augmented and Virtual Reality}
Inertial sensors are pivotal in augmenting user interaction within AR and VR technologies through enhanced movement tracking and prediction capabilities. The HOOV system demonstrates this by using wrist-worn inertial sensors to track hand positions outside the user's visual field, improving interaction with virtual objects and spatial awareness \cite{streli2023hoov}. This concept is further developed in studies like \cite{kahanowich2023learning} and \cite{bao2023uncertainty}, where inertial sensors facilitate real-time human arm movement prediction for effective human-robot collaboration and enhance 3D hand trajectory forecasting in VR by integrating inertial data with visual inputs, respectively. Extending these applications, \cite{wang2017optimal}, and \cite{zhang2022rids} explore the use of inertial sensors in generating collision-free robot trajectories, predicting complex upper limb movements, and improving gesture recognition algorithms in VR, significantly enhancing both safety and efficiency in human-robot interactions.

Moreover, \cite{gamage2021so} explores the continuous 3D hand trajectory prediction in VR, highlighting the potential of kinematics-based models to predict user interactions with virtual objects, thus preventing collisions and enhancing user experience. Finally, \cite{viola2021trace} discuss the use of motion tracking and prediction in enhancing the streaming of content in AR applications, where accurate prediction of the user's field of view can significantly optimize bandwidth usage and reduce latency.
\section{Implementation}
Our focus is on the inertial sensor present in the smartphone. We operate under the assumption that IMU data from the smartphone is always available. Utilizing historical IMU data, we predict the future position of the hand holding the smartphone. The IMU is known to consume less power than the camera sensor, making it more suitable for continuous sensing without causing significant resource consumption on the device.

\subsection{Model}
For the learning architecture, we use a two-layer Bidirectional LSTM with exogenous input, inspired by prior works \cite{yi2021transpose, mollyn2023imuposer}. For the available IMU, our system uses historical data of orientation (represented as a quaternion) as well as acceleration as input, both in a global coordinate frame of reference. We then flatten the historical data of hand positions concatenated with historical IMU data of $n$ sequence length to create an model input vector of size $n \times 10$: 3 historical hand position, 4 orientation, and 3 acceleration. We then create a exogenous input (dimension=7) to model of orientation and a acceleration at timestep $n+1$, as we assumed IMU data is always available. With these input model predict future hand position at time step $n + 1$.

The LSTM layer is central to handling the sequential data, with its bidirectional configuration allowing the model to learn dependencies from both forward and backward sequences. This LSTM layer consists of 2 layers and a hidden dimension of 256 for each direction. Thus, the bidirectional setup effectively doubles the LSTM output features to 512 per sequence. After processing through the LSTM, the output at the last time step, representing the most recent and relevant features from both directions of the sequence, is concatenated with the exogenous sensor data, resulting in a combined input vector of 519 dimensions (512 from LSTM and 7 from the exogenous data). This combined data is first passed through a fully connected layer with a dimension transformation from 519 to 256, integrating the features using a ReLU activation function for non-linear processing. A dropout layer with a rate of 0.2 follows to prevent overfitting by randomly dropping units during the training phase. Finally, the output is passed through another fully connected layer, which reduces the dimension from 256 to 3, corresponding to the three dimensions of the hand position.
\begin{figure*}[h]
  \centering
  \includegraphics[width=\linewidth]{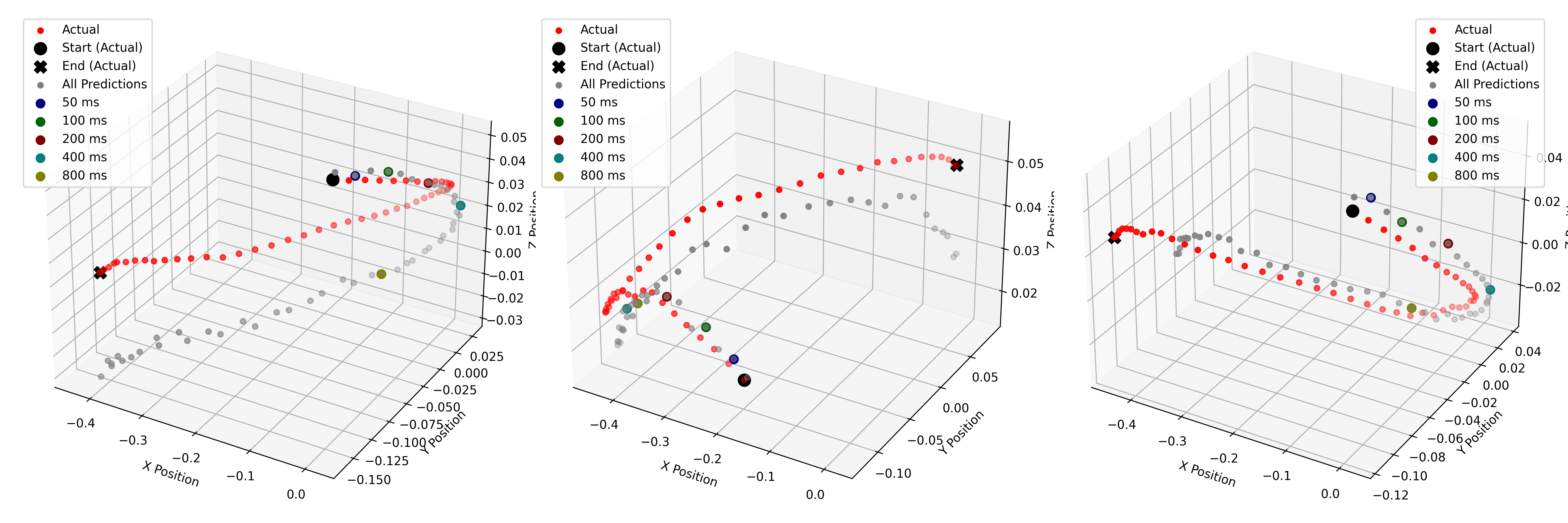}
  \caption{Samples of predictions by our model at 50, 100, 200, 400, and 800 ms.}
  \label{fig:yourlabel}
\end{figure*}

\begin{figure}[h]
    \centering
    \begin{subfigure}{0.35\textwidth}
        \centering
        \includegraphics[width=\linewidth]{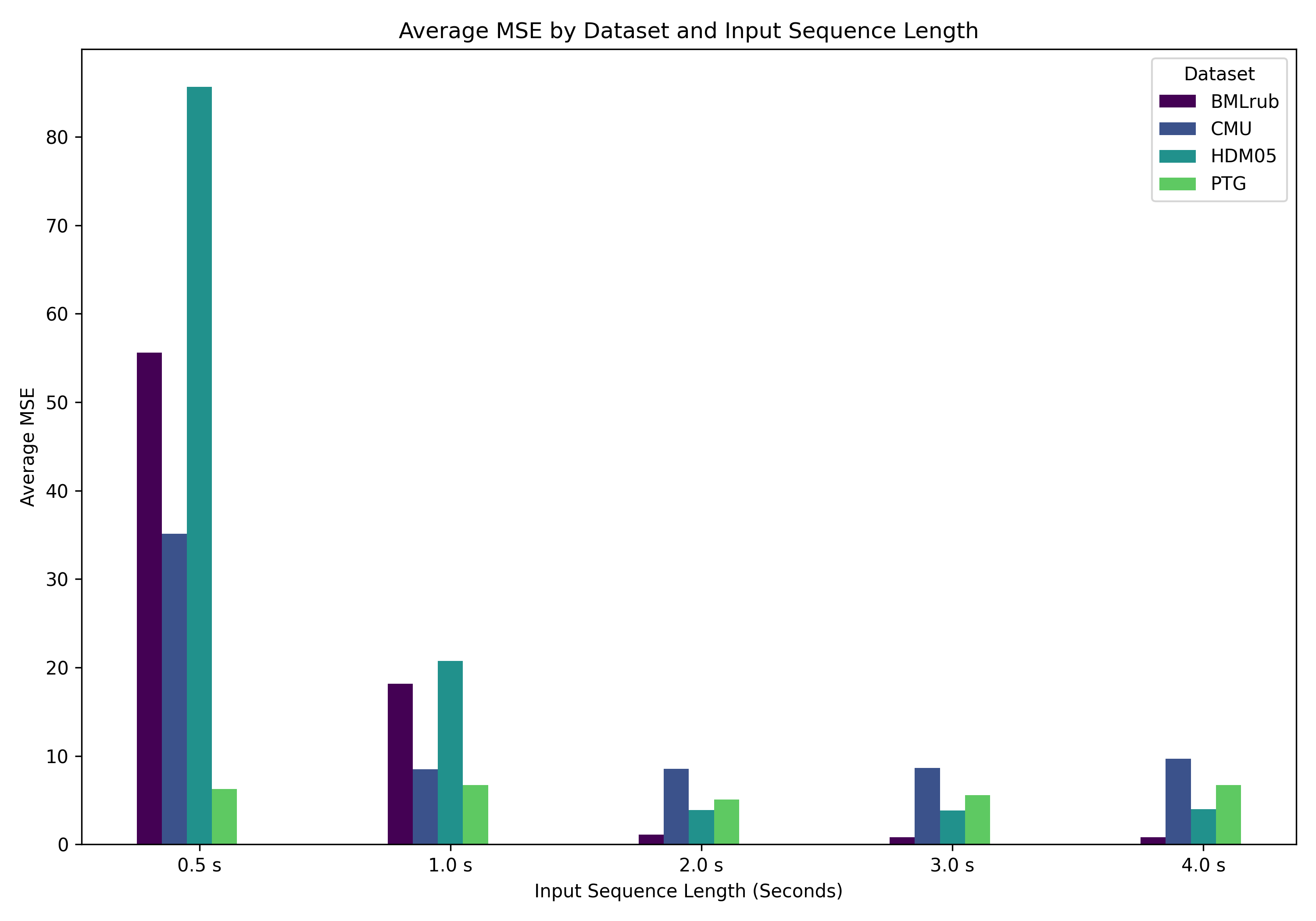}
        \caption{Average MSE across time horizons by dataset and input sequence length}
        \label{fig:mae_input_threshold}
    \end{subfigure}
    \hfill
    \begin{subfigure}{0.35\textwidth}
        \centering
        \includegraphics[width=\linewidth]{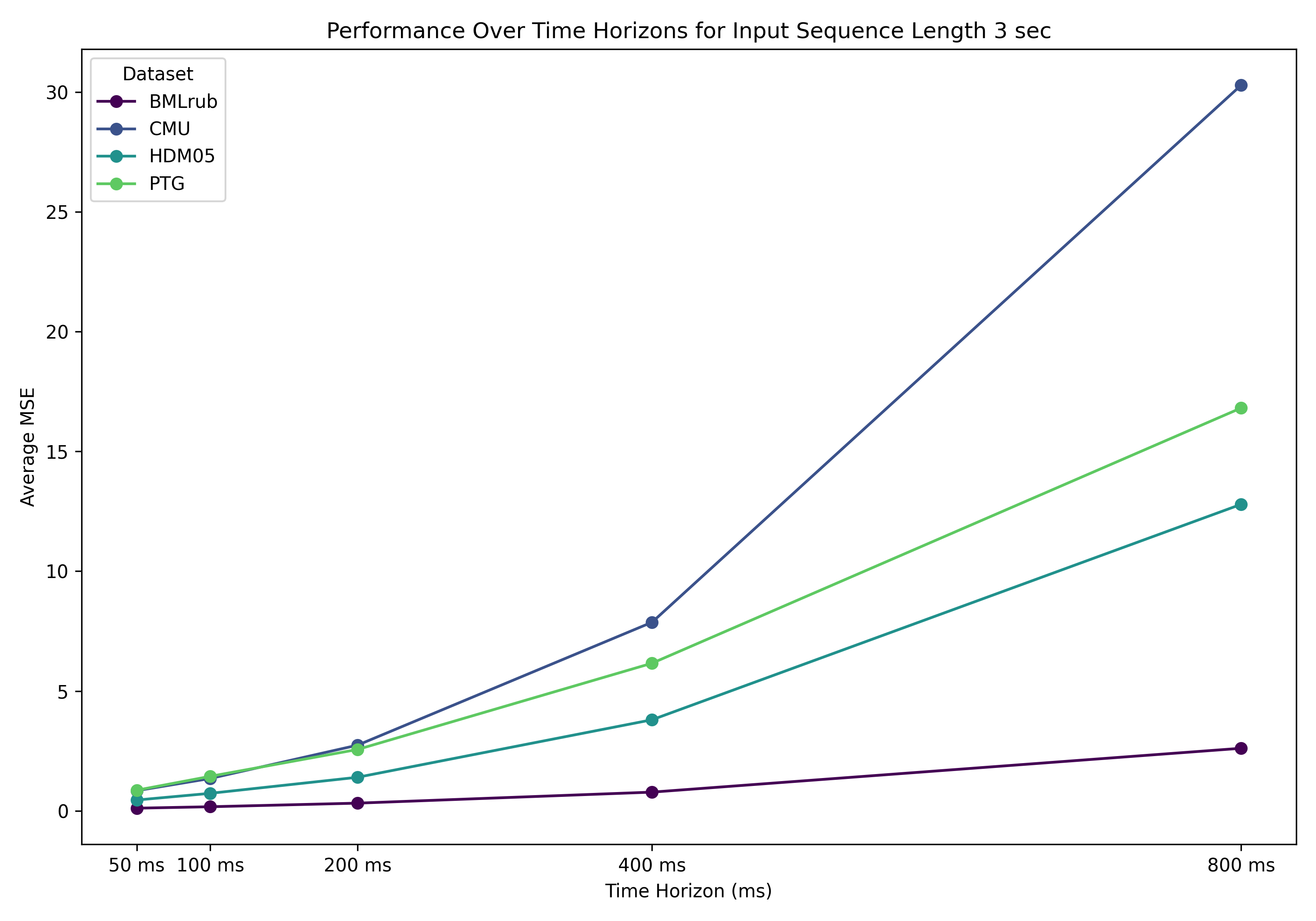}
        \caption{Performance over time horizons for input sequence length 3 sec}
        \label{fig:performance_degradation}
    \end{subfigure}
    \caption{Comparison of average MAE across different datasets and time horizons.}
    \label{fig:comparison_figures}
\end{figure}

\subsection{IMU Dataset Synthesis} \label{IMUDS}
We required a significant volume of data to train our future hand position model. We employ the CMU \cite{mocap}, BMLrub \cite{troje2002decomposing}, and HDM05 \cite{muller2007documentation} subsets from the AMASS \cite{mahmood2019amass} dataset for the training and testing of our hand position model. The AMASS dataset aggregates various optical marker-based MoCap datasets and standardizes them into 3D human meshes through the SMPL \cite{loper2023smpl} model parameters, creating a comprehensive human motion database. It's important to note that AMASS has been used in several prior studies  \cite{huang2018deep, yi2021transpose} as the foundation for creating synthetic datasets.

We utilize the synthetic data creation method as presented in TransPose \cite{yi2021transpose} and DIP \cite{huang2018deep}. Essentially, we affix virtual IMUs to particular vertices in the SMPL mesh at the right wrist and generate synthetic acceleration data from neighboring frames in the global reference frame. For the creation of synthetic orientation data, we compute joint rotations in relation to the global frame by compounding local rotations from the joint towards the pelvis (root), adhering to the SMPL kinematic chain.

\subsection{Training}
The model is trained end-to-end using the PyTorch and PyTorch Lightning deep learning frameworks. The batch size is set to 64 during training, and the Adam optimizer is utilized to update the weights with a learning rate of 0.00001. This process is guided by a learning rate scheduler set to plateau. Training uses non-overlapping 3-second windows (or 180 sequence length) of paired IMU and translation data. The model is trained to predict future hand positions using the mean squared error (MSE) loss. The entire training, lasting 100 epochs or approximately 34 hours, is conducted on an NVIDIA Tesla V100.

\section{Evaluation}
\subsection{Dataset}We evaluate our method subset (CMU \cite{mocap} (9.19hr), BMLrub \cite{troje2002decomposing} (1.7hr), and HDM05 \cite{muller2007documentation} (2.4hr))  of AMASS dataset and on smartphone based full body pose estimatimate data - Pose-on-the-Go (PTG) \cite{ahuja2021pose}. It's worth noting that PTG is collected in the real world. Evaluating our method using PTG further demonstrates its applicability in real-world scenarios. We synthesis IMU and hand position translation on AMASS to train and test our model (See Section \ref{IMUDS}). Furthermore the PTG dataset already have hand position (smartphone position) and orientation data. We compute the acceleration with the help of hand position translation and timestamp in the dataset for further training and evaluation. 

\subsection{Evaluation Results}

We evaluated four different datasets in total. To further assess the generalization capability of our proposed method, we conducted a 4-fold cross-dataset evaluation. This involved training on three subsets and testing on the remaining subset in a round-robin fashion. Table \ref{tab:cross_dataset_eval} displays the experimental results of various methods tested on the CMU, BMLrub, HDM05, and PTG datasets. To evaluate our model for future hand position prediction, we try non-overlapping 0.5, 1, 2, 3, and 4-second windows of paired IMU and translation data, i.e., $n$ (sequence length) and IMU data at $n+1$ time step, to predict hand position at $n+1$ time step, we assume that IMU data is always available. We use the translation predicted in $n+1$ time step as input to predict future position at $n+2$, and so on. We report the MSE score for each dataset at a prediction horizon of 50, 100, 200, 400, and 800 ms at 30fps.

\begin{table}[h]
\centering
\caption{Results of cross-dataset evaluation for input sequence length 3 sec. The table shows average MSE in mm.}
\label{tab:cross_dataset_eval}
\begin{tabular}{lrrrr}
\toprule
 Time Horizon (ms) &    CMU &  BMLrub &  HDM05 &     PTG \\
\midrule
                50 &   0.84 &    0.11 &   0.45 &    5.36 \\
               100 &   1.35 &    0.17 &   0.73 &    9.32 \\
               200 &   2.74 &    0.32 &   1.40 &   19.66 \\
               400 &   7.87 &    0.78 &   3.80 &   52.74 \\
               800 &  30.28 &    2.61 &  12.79 &  143.62 \\
\bottomrule
\end{tabular}
\end{table}

Our results (Fig \ref{fig:comparison_figures}) show that all prediction errors stopped reducing after 3 seconds of input data. The errors increase (Table \ref{tab:cross_dataset_eval} and Fig \ref{fig:performance_degradation}) as the prediction time horizon increases across all datasets. This is expected as longer prediction times generally introduce more uncertainty into the estimation process. The increasing trend in error rates at longer time horizons suggests a limit to the predictability of movement using MotionTrace, especially for datasets with complex movement dynamics like CMU and PTG. For effective AR applications, focusing on shorter prediction windows or improving the model's ability to handle complex movements might be necessary.

\section{Discussion and Conclusion}
FOV prediction is essential for enhancing the interactivity and immersion of smartphone AR experiences. To this end, we propose MotionTrace, a method to continuously estimate a user's hand position in 3D-space. This enables phone position localization for FOV prediction using inertial sensors. We evaluated our method on a large motion capture (AMASS) and smartphone-based full body pose estimation (Pose-on-the-Go) dataset. Our method was able to predict future hand positions with an average MSE ranging from 0.11 - 143.62 mm across different datasets for time horizons between 50 - 800 ms. We also found that 3 seconds of historical inertial sensor data is sufficient for making this prediction. However, our results showed that errors increase as the prediction time horizon lengthens, leading to larger errors across all datasets. The best results were found within a prediction time horizon of 50 - 400 ms.

Our method, when used in conjunction with other methods proposed in previous literature \cite{viola2021trace, wang2020towards, van2019towards, park2018volumetric}, can provide incremental utility. We base our work on the premise that inertial sensor data is always available on the phone. It's also presumed to consume less power than the camera sensor, making it more suitable for continuous sensing without significantly straining the device's resources. This presents a significant advantage over previous methods \cite{viola2021trace, wang2020towards} that rely on continuous historical streaming to predict the FOV.






%



\bibliographystyle{IEEEtran}
\bibliography{Reference}

\end{document}